\documentclass[aps,12pt,prd,preprintnumbers,nofootinbib]{revtex4}

\usepackage{amsmath}
\usepackage{amsfonts}

\newcommand{\pop}[1]{\frac{\partial}{\partial #1}}

\newcommand{\p}[1]{(\ref{#1})}
\newcommand{\lb}{\label}
\newcommand{\ca}[1]{{\cal #1}}

\newcommand{\be}{\begin{eqnarray}}
\newcommand{\ee}{\end{eqnarray}}
\newcommand{\nn}{\nonumber\\}

\newcommand{\F}{{\cal F}}
\newcommand{\A}{{\cal A}}

\newcommand{\U}{{\rm U}}
\newcommand{\SU}{{\rm SU}}
\newcommand{\SO}{{\rm SO}}

\newcommand{\blanc}{\ \ \ \ \ \ \ }

\newcommand{\N}{{\cal N}}

\newcommand{\vp}{\varphi}
\newcommand{\ve}{\varepsilon}

\begin{document}
\begin{flushright}
\preprint{ITEP-TH-19/10}
\end{flushright}

\title{${\cal N}=4$, $3D$ Supersymmetric Quantum Mechanics\break in Non-Abelian Monopole Background}

\author{{\bf  Evgeny Ivanov}}
\affiliation{Bogoliubov Laboratory of Theoretical Physics, JINR, 141980 Dubna, Russia}
\email{eivanov@theor.jinr.ru}

\author{{\bf Maxim Konyushikhin}}
\affiliation{SUBATECH, Universit\'e de Nantes, 4 rue Alfred Kastler, BP 20722, Nantes 44307, France}
\affiliation{Institute for Theoretical and Experimental Physics,B. Cheremushkinskaya 25, Moscow 117259, Russia}
\email{konush@itep.ru}

\begin{abstract}
\vskip 1cm\ \noindent Using the harmonic superspace approach, we
construct the three-dimensional ${\cal N}=4$
supersymmetric quantum mechanics of the supermultiplet $({\bf 3, 4,
1})$ coupled to an external SU(2) gauge field. The off-shell ${\cal
N}=4$ supersymmetry requires the gauge field to be a
static form of the 't Hooft ansatz for the $4D$ self-dual SU(2)
gauge fields, that is a particular solution of Bogomolny equations
for BPS monopoles. We present the explicit form of the corresponding
superfield and component actions, as well as of the quantum
Hamiltonian and ${\cal N}=4$ supercharges. The latter can be used to
describe a more general ${\cal N}=4$ mechanics system, with an
arbitrary BPS monopole background and on-shell ${\cal N}=4$
supersymmetry. The essential feature of our construction is the use
of semi-dynamical spin $({\bf 4, 4, 0})$ multiplet with the
Wess-Zumino type action.

\end{abstract}


\maketitle

\section{Introduction}
The models of supersymmetric quantum mechanics (SQM) with background gauge fields
are of obvious interest for a few reasons. One reason is the close relation of these systems
to the renowned Landau problem and its generalizations (see e.g.~\cite{Land}).
The Landau-type models constitute a basis of the theoretical description
of quantum Hall effect (QHE), and it is natural to expect that their supersymmetric
extensions, with extra fermionic variables added, may be relevant to
spin versions of QHE. Also, these systems can provide quantum-mechanical realizations
of various Hopf maps closely related to higher-dimensional QHE (see e.g. \cite{Hopf} and references therein).
At last, they exhibit $d=1$ prototypes of couplings to higher-$p$ forms in superbranes
and so offer a simplified framework to study these couplings.

${\cal N}=4$ SQM models with the background Abelian gauge fields were treated in the pioneer
papers \cite{de Crombrugghe:1982un,Smilga:1986rb} and, more recently, e.g. in \cite{IKL,IvLecht,Kirchberg:2004za,KonSmi}.
In particular, in \cite{IvLecht} an off-shell Lagrangian superfield formulation
of the general models associated with
the multiplets $({\bf 4, 4, 0})$ and $({\bf 3, 4, 1})$ was given
in the ${\cal N}=4, d=1$ harmonic superspace.\footnote{The first
superfield formulation of general $({\bf 3, 4, 1})$ SQM
(without background gauge field couplings) was given in \cite{IvSmi}.}
It was found that ${\cal N}=4, d=1$ supersymmetry
requires the gauge field to be self-dual in the four-dimensional $({\bf 4, 4, 0})$ case,
or to obey a ``static'' version of the self-duality condition in  the three-dimensional $({\bf 3, 4, 1})$ case.
In the papers \cite{Kirchberg:2004za,KonSmi} it was observed (in a Hamiltonian approach) that the Abelian $({\bf 4, 4, 0})$ ${\cal N}=4$ SQM
admits a simple generalization to arbitrary self-dual non-Abelian background.\footnote{The presence of ${\cal N}=4$
supersymmetry in the Dirac operator with a self-dual gauge field was established first in \cite{JR},
though in an implicit way.}
In \cite{Ivanov:2009tw} an off-shell Lagrangian
formulation was shown to exist for a particular class of such non-Abelian ${\cal N}=4$ SQM models, with $\SU(2)$ gauge group
and 't~Hooft ansatz \cite{tHooft} for the self-dual $\SU(2)$ gauge field (see also \cite{KLS}). As in the Abelian case, it was the use of ${\cal N}=4, d=1$
harmonic superspace that allowed us to construct such an off-shell formulation.
A new non-trivial feature of the construction of \cite{Ivanov:2009tw} is the involvement of
an auxiliary ``semi-dynamical'' $({\bf 4, 4, 0})$ multiplet with the Wess-Zumino
type action possessing an extra  gauged $\U(1)$ symmetry. After quantization, the corresponding bosonic $d=1$ fields become
a sort of spin $\SU(2)$ variables to which the background gauge field naturally couples.\footnote{The use of such
auxiliary bosonic variables for setting up coupling of a particle to Yang-Mills fields  can be traced back to \cite{Bala}.
In the context of ${\cal N}=4$ SQM, they were employed in \cite{FIL0,FIL} and \cite{Hopf,BKS}.}

In the present paper, we exploit a similar method to construct ${\cal N}=4$ supersymmetric coupling
of the multiplet $({\bf 3, 4, 1})$ to an external non-Abelian gauge field. Like in the $({\bf 4, 4, 0})$ case,
it is the $d=1$ harmonic superspace which makes it possible to perform such a construction in a general form.
Off-shell ${\cal N}=4$ supersymmetry is shown to
restrict the external gauge field to a ``static'' version of the 't Hooft ansatz for four-dimensional
self-dual $\SU(2)$ gauge field, that is to a particular solution of the general monopole Bogomolny
equations \cite{Bog}.\footnote{Some BPS monopole backgrounds in the framework of ${\cal N}=2$ SQM were considered, e.g., in \cite{H1}.}
A new feature of the $({\bf 3, 4, 1})$ case is the appearance of ``induced'' potential term in the on-shell action
as a result of eliminating the auxiliary field of the $({\bf 3, 4, 1})$ multiplet. This term is bilinear in the
$\SU(2)$ gauge group generators. As a particular ``spherically symmetric''  case of our construction (with the
exact $\SU(2)$ R-symmetry) we recover, up to an essentially different treatment of the spin variables,
the ${\cal N}=4$ mechanics with Wu-Yang monopole \cite{WYa} recently considered
in \cite{BKS}.

\section{Superfield Formulation}

In the ${\cal N}=4$, $d=1$ harmonic superspace (HSS) approach \cite{IvLecht}, the superfields depend on bosonic variables $t,\  u^{\pm \alpha}$,
where the harmonics $u^{+\alpha}$, $u^{-}_{\alpha} = (u^{+\alpha})^*$, $u^{+\alpha} u^-_\alpha = 1$ parametrize the R-symmetry
group $\SU(2)$ of the ${\cal N} = 4$ superalgebra, and on fermionic
variables $\theta^\pm =  \theta^\alpha u_\alpha ^\pm$,
$\bar\theta^\pm = \bar\theta^\alpha u_\alpha ^\pm $. The most
important feature of HSS is the presence of an {\sl analytic
subspace} $\left\{t_{\rm A}, \theta^+, \bar\theta^+,
u^\pm_\alpha\right\}$ in it involving the  ``analytic time'' $t_{\rm A} = t + i(\theta^+
\bar \theta^- + \theta^- \bar \theta^+)$ and containing twice as
less fermionic coordinates.
Spinor derivatives $D^+$ and $\bar D^+$ in the {\sl analytic basis}
$\left\{t_{\rm A},\theta^\pm,\bar\theta^\pm, u_\alpha^\pm\right\}$ are \cite{HSS}
\begin{equation}\label{Dana}
  D^+=\pop{\theta^-},\blanc \bar D^+=-\pop{\bar\theta^-}.
\end{equation}
Other important objects used in what follows are the harmonic derivatives $D^{++}$, $D^{--}$ preserving
the ${\cal N}=4$ analyticity:
\begin{eqnarray}
  D^{++}=u^+_\alpha\pop{u^-_\alpha}+\theta^+\pop{\theta^-}+\bar\theta^+\pop{\bar\theta^-}
    +2i\theta^+\bar\theta^+ \pop{t_{\rm A}},
\\
  D^{--}=u^-_\alpha\pop{u^+_\alpha}+\theta^-\pop{\theta^+}+\bar\theta^-\pop{\bar\theta^+}
    +2i\theta^-\bar\theta^- \pop{t_{\rm A}} .
\end{eqnarray}
Also, for further use, we give how the coordinates of the analytic subspace transform under ${\cal N}=4$ supersymmetry:
\be
\delta \theta^+ = \epsilon^\alpha u^+_\alpha\,, \quad \delta \bar\theta^+ = \bar\epsilon^\alpha u^+_\alpha\,, \quad
\delta t_A = 2i\left(\epsilon^\alpha u^-_\alpha \bar\theta^+ - \bar\epsilon^\alpha u^-_\alpha\theta^+ \right), \quad \delta u^\pm_\alpha = 0\,,
\;\; \bar\epsilon^\alpha
= (\epsilon_\alpha)^*\,. \lb{N4transf}
\ee

In this paper, we shall deal with the analytic superfields $L^{++}$ and $v^+, \widetilde{v^+}$ which encompass, respectively,
the multiplets $(\bf{3,4,1})$ and $(\bf{4,4,0})$ and are subjected to the constraints
\begin{equation}\label{a}
\begin{array}{l}
\mbox{(a)}\ D^{+}L^{++} = \bar D^{+}L^{++} = 0\,,
\blanc\;\;\;\;\;\;\;\,
 \mbox{(b)}\ D^{++}L^{++} = 0\,,\;\;
 \widetilde{(L^{++})} = -L^{++},\blanc\blanc\,
\end{array}
\end{equation}
\begin{equation}\label{b}
\begin{array}{l}
\mbox{(a)}\ D^{+}(v^{+}, \;\widetilde{v^{+}}) = \bar D^{+} (v^{+}, \;\widetilde{v^{+}}) =0\,,\;\;
 \,\mbox{(b)}\ (D^{++} + i V^{++}) v^{+} = (D^{++} - i V^{++}) \widetilde{v^{+}} = 0\,.
\end{array}
\end{equation}
The $\U(1)$ gauge superfield $V^{++}$ appearing in Eqs. (\ref{b}b) is analytic,
\begin{equation}
\label{constrV}
 D^+ V^{++}  = \bar D^+ V^{++} = 0 \, ,
\end{equation}
and pseudoreal, $V^{++} = \widetilde{(V^{++})}.\,$ It ensures the covariance of (\ref{b}b) under
the gauge $\U(1)$ transformations with the analytic parameter $\Lambda$ \cite{DI}
\begin{equation}
\label{gaugeharm}
V^{++} \ \to \ V^{++} + D^{++} \Lambda,\blanc
 v^+ \ \to\ e^{-i\Lambda} v^+ ,\;\;  \widetilde{v^+} \ \to\ e^{i\Lambda} \widetilde{v^+} ,
\quad
D^+\Lambda=\bar D^+\Lambda=0\ .
\end{equation}
In what follows, we shall use the WZ gauge for $V^{++}$,
\begin{equation}
\label{VPP}
V^{++} \ =\ 2i\, \theta^+ \bar \theta^+ B\,.
\end{equation}
Here $B(t)$ is a real $d=1$ ``gauge field'',
it transforms as $B \ \to \ B + \dot\lambda\,$, with $\lambda(t)$ being the parameter
of the residual gauge U(1) symmetry.

The constraints (\ref{a}a), (\ref{b}a)  and \p{constrV}
are the ${\cal N}=4$ Grassmann analyticity conditions just implying that the superfields
$L^{++}$, $v^+$, $\widetilde{v}^+$, $V^{++}$ live
on the analytic superspace $\left\{t_{\rm A}, \theta^+, \bar\theta^+, u^\pm_\alpha\right\}\,$.
The basic conditions are those with the harmonic derivatives, i.e. (\ref{a}b) and (\ref{b}b).
They constrain the analytic superfields $L^{++}$ and $v^+\,,$ $\widetilde{v^+}$ to have the appropriate off-shell component field contents, namely
$({\bf 3, 4, 1})$ and $({\bf 4, 4, 0})$:
\begin{equation}
L^{++} = \ell^{\alpha\beta}u^+_\alpha u^+_\beta
+ i\theta^+ \chi^\alpha u^+_\alpha
+i\bar\theta^+ \bar\chi^\alpha u^+_\alpha +
\theta^+\bar\theta^+ [F - 2 i\dot{\ell}^{\alpha\beta} u^+_{\alpha} u^-_{\beta}],
\end{equation}
with $\left(\ell_{\alpha\beta}\right)^* = -\ell^{\alpha\beta}\,, \; (\chi^\alpha)^* = \bar\chi_\alpha\,$,
and
\be
&& v^+=\phi^\alpha u^+_\alpha
  + \theta^+\omega_1 + \bar\theta^+\bar\omega_2
  -2i\theta^+\bar\theta^+  (\dot \phi^\alpha + i B\phi^\alpha) u^-_\alpha\ ,\label{vp}\\
&& \widetilde  {v^+}= \bar \phi^\alpha u^+_\alpha
  + \theta^+\omega_2  -\bar\theta^+ \bar\omega_1
  -2i\theta^+\bar\theta^+ ({\dot {\bar \phi}^\alpha} - iB \bar \phi^\alpha) u^-_\alpha\,, \label{vptild}
\ee
with $\bar \phi^\alpha = (\phi_\alpha)^*$, $\bar\omega_{1,2}=(\omega_{1,2})^*$.
The multiplet $L^{++}$ involves the 3-dimensional target space coordinates $\ell^{\alpha\beta} = \ell^{\beta\alpha}$, their
fermionic partners and a real auxiliary field $F$, while $v^+$ accommodates the auxiliary degrees of freedom needed
to arrange a coupling to the external non-Abelian $\SU(2)$ Yang-Mills field \cite{Ivanov:2009tw}.

The full Lagrangian ${\cal L}$ entering the  $\N=4$ invariant  off-shell action $S= \int dt {\cal L}$ consists of the three pieces
\begin{multline}
{\cal L} = {\cal L}_{\rm kin} + {\cal L}_{\rm int} + {\cal L}_{\rm FI}
  = \int du\, d^4\theta\, R_{\rm kin}(L^{++}, L^{+-}, L^{--}, u)
\\[3mm]
- \frac{1}{2} \int du\, d\bar\theta^+ d\theta^+\, K(L^{++}, u) v^+\widetilde{v^+}
-\frac {i k}2  \int  \, du \, d\bar \theta^+ d\theta^+ \, V^{++}
\,, \lb{ACT}
\end{multline}
where $L^{+-}=\frac 12D^{--}L^{++}$ and $L^{--}=D^{--}L^{+-}$. The superfield functions $R_{\rm kin}$ and $K$
bear an arbitrary dependence on their arguments. The meaning of three terms in \p{ACT} will be explained in the next
Section.

\section{From harmonic superspace to components}

The first, sigma-model-type term in Eq.~\p{ACT}, after integrating over Grassmann and harmonic variables, yields the generalized
kinetic terms for $\ell^{\alpha\beta}, \chi^\alpha, \bar\chi_\alpha$:
\begin{multline}\label{kin_term}
{\cal L}_{\rm kin}
  =\frac 18f^{-2}\left(-2\dot \ell_{\alpha\beta}\dot\ell^{\alpha\beta}+F^2\right)
  +\frac i8 f^{-2}\left(\bar\chi_\alpha\dot\chi^\alpha-{\dot{\bar\chi}}_\alpha\chi^\alpha\right)
  +\frac 1{64}\left(\partial_{\alpha\beta}\partial^{\alpha\beta}f^{-2}\right)\chi^4
\\[2mm]
  +\frac{i}{4f^3}\dot \ell^{\alpha\beta}\big\{
    \partial_{\alpha\gamma}f\chi_\beta\bar\chi^\gamma+\partial_{\beta\gamma}f\chi^\gamma\bar\chi_\alpha
  \big\}
  -\frac{1}{4f^3}F\chi^\alpha\bar\chi^\beta\partial_{\alpha\beta}f,
\end{multline}
where $\chi^4=\chi^\alpha\chi_\alpha \bar\chi^\beta\bar\chi_\beta$,
$\partial_{\alpha\beta} \equiv \frac{\partial}{\partial \ell^{\alpha\beta}}$
and $f(\ell)$ is a conformal factor.\footnote{
The calculations are most easy in the central basis, where
$L^{++}=u^+_\alpha u^+_\beta L^{\alpha\beta}\left(t,\theta_\gamma,\bar\theta^\delta\right)$. Then
\begin{equation*}
f^{-2}(\ell)=-\partial_{\alpha\beta}\partial^{\alpha\beta}
  \int R_{\rm kin}\left(
    \ell^{\alpha\beta}u^+_\alpha u^+_\beta,
    \ell^{\alpha\beta}u^+_\alpha u^-_\beta,
    \ell^{\alpha\beta}u^-_\alpha u^-_\beta
  \right)\,du.
\end{equation*}
}
The fermionic kinetic term can be brought to the canonical form by the change of variables
\begin{equation}\label{changetopsi}
 \chi^\alpha=2f \psi^\alpha,\blanc
 \bar\chi_\alpha=2f\bar\psi_\alpha.
\end{equation}
It is worth pointing out that the R-symmetry $\SU(2)$ group amounts to the rotational ${\rm SO(3)}$ group in the $\mathbb{R}^3$ target space
parametrized by $\ell^{\alpha\beta}$. The conformal factor $f(\ell)$ can bear an arbitrary dependence on $\ell^{\alpha\beta}$,
so this ${\rm SO(3)}$ can be totally broken in the Lagrangian \p{kin_term}.

The second piece in Eq.~(\ref{ACT}) describes the coupling to an external non-Abelian gauge field.
Performing the integration over $\theta^+$, $\bar\theta^+$ and $u^\pm_\alpha$, eliminating the auxiliary fermionic fields $\omega_{1,2}$
and, finally, rescaling the bosonic doublet variables as $\varphi_\alpha \ =\ \phi_\alpha \sqrt {h(\ell) }$, where
\begin{equation}
 h(\ell)=\int du\, K\left(\ell^{\alpha\beta}u^+_\alpha u^+_\beta,u^\pm_\gamma\right), \lb{hdef}
\end{equation}
after some algebra we obtain
\begin{equation}\label{spinor_lagr}
{\cal L}_{\rm int}=i\bar\vp^\alpha\left(\dot\vp_\alpha+iB\vp_\alpha\right)
  +\bar\vp^\gamma\vp^\delta \frac 12
    \left(\A_{\alpha\beta}\right)_{\gamma\delta} \dot \ell^{\alpha\beta}
  -\frac 12 F\,\bar\vp^\gamma\vp^\delta \,U_{\gamma\delta}
    +\frac 14 \chi^\alpha\bar\chi^\beta\bar\vp^\gamma\vp^\delta
      \nabla_{\alpha\beta}U_{\gamma\delta}.
\end{equation}
Here the non-Abelian background gauge field and the scalar (matrix) potential are fully specified by the function $h$ defined in \p{hdef}:
\begin{equation}
\left(\A_{\alpha\beta}\right)_{\gamma\delta}
  =\frac i{2h} \Big\{
    \ve_{\gamma\beta}\partial_{\alpha\delta}h
    +\ve_{\gamma\alpha}\partial_{\beta\delta}h
    +\ve_{\delta\beta}\partial_{\alpha\gamma}h
    +\ve_{\delta\alpha}\partial_{\beta\gamma}h
  \Big\},\blanc
 U_{\gamma\delta}=\frac{1}{h}\partial_{\gamma\delta}h\,. \lb{GpotMpot}
\end{equation}
By definition, the function $h$ obeys the 3-dimensional Laplace equation,
\begin{equation}
\partial^{\alpha\beta}\partial_{\alpha\beta}\,h = 0\,. \lb{Lapl}
\end{equation}

Using the explicit expressions \p{GpotMpot}, it is straightforward to check the relation
\begin{equation}\label{samo}
 \left(\F_{\alpha\beta}\right)_{\gamma\delta}
   =2i\nabla_{\alpha\beta}U_{\gamma\delta},
\end{equation}
where
\begin{equation}\label{fij}
 \left(\F_{\alpha\beta}\right)_{\gamma\delta}
 =-2\partial_{\alpha}^{\;\lambda}\left(\A_{\lambda\beta}\right)_{\gamma\delta}
  +i\left(\A_{\alpha}^{\,\,\,\lambda}\right)_{\!\gamma\sigma}
    \left(\A_{\lambda\beta}\right)^{\,\,\sigma}_{\delta}
  +\left(\alpha\leftrightarrow \beta\right),
\end{equation}
\begin{equation}\label{nabla}
 \nabla_{\alpha\beta}U_{\gamma\delta}=-2\partial_{\alpha\beta}U_{\gamma\delta}
  +i\left(\A_{\alpha\beta}\right)_{\gamma\lambda}U^{\,\,\lambda}_{\!\delta}
  +i\left(\A_{\alpha\beta}\right)_{\delta\lambda}U^{\,\,\lambda}_{\!\gamma},
\end{equation}
and $\left(\F_{\alpha\beta}\right)_{\gamma\delta}$ is related to the standard gauge field strength in the vector notation, see below.
As we shall see soon, the condition \p{samo} is none other than the static
form of the general self-duality condition for the $\SU(2)$ Yang-Mills field on $\mathbb{R}^4\,$, i.e.
the Bogomolny equations for BPS monopoles \cite{Bog},
while \p{GpotMpot} provides a particular solution to these equations, being a static form
of the renowned 't Hooft ansatz \cite{tHooft}.

Note that the relation \p{samo} is covariant and the Lagrangian \p{spinor_lagr} is form-invariant under the following
``target space'' $\SU(2)$ gauge transformations:
\begin{equation}\label{grep}
\begin{array}{c}
 \varphi_\alpha\rightarrow\left(U^\dagger\varphi\right)_\alpha,\,\,\,\,
 \bar\varphi^\alpha\rightarrow \left(\bar\varphi U\right)^\alpha
\\[2mm]
 \A_{\alpha\beta}\rightarrow \Lambda^\dagger \A_{\alpha\beta} \Lambda+i\Lambda^\dagger\partial_{\alpha\beta} \Lambda,\blanc
 U\rightarrow \Lambda^\dagger U \Lambda,
\end{array}
\end{equation}
with $\Lambda(\ell)\in \SU(2)$.
This is not a genuine symmetry; rather, it is a reparametrization of the Lagrangian which allows one to cast
the background potentials \p{GpotMpot}  in some different equivalent forms.
It is worth noting that the gauge group indices coincide with those of the R-symmetry group, like in the
four-dimensional case \cite{Ivanov:2009tw}. Nevertheless, the ``gauge'' reparametrizations \p{grep} do not affect
the doublet indices of the target space coordinates $\ell^{\alpha\beta}$ and their superpartners
accommodated by the superfield $L^{++}$. They act only on the semi-dynamical spin variables
$\varphi_\alpha, \bar\varphi^\alpha$ and gauge and scalar potentials \p{GpotMpot}.

Finally, the last piece in Eq.~(\ref{ACT}) yields the Fayet-Iliopoulos term,
\begin{equation}\label{FI}
 {\cal L}_{\rm FI}=k B\,.
\end{equation}
In the quantum case, the coefficient $k$ is quantized, $k \in \mathbb{Z}\,$, on the same ground as in the 4-dimensional case \cite{Ivanov:2009tw}.
As is obvious from Eqs.~(\ref{spinor_lagr}) and (\ref{FI}), the auxiliary gauge field $B$ serves as a Lagrange multiplier for the constraint
\be
\bar\varphi^\alpha \varphi_\alpha = k\,.\lb{Cconstr}
\ee
In the classical case it implies (together with the residual $\U(1)$ gauge freedom) that $\bar\varphi^\alpha, \varphi_\alpha$
describe coordinates on a sphere ${\rm S^2}$ in the target space, while in the quantum case the constraint (\ref{Cconstr}) is imposed on the wave function
requiring it to span an irreducible $\SU(2)$ multiplet with spin $|k|/2$ \cite{Ivanov:2009tw}.

It is instructive to rewrite the above relations and expressions, including the full Lagrangian (\ref{ACT})
in a vector notation. To this end, we associate a vector $v_i$ to any traceless bi-spinor $v_\alpha^\beta$ by the general rule
\begin{equation}
 v_\alpha^\beta=v_i\left(\sigma_i\right)_{\!\alpha}^{\,\,\beta},\blanc
  v_i=\frac 12 v^\alpha_\beta\left(\sigma_i\right)_{\!\alpha}^{\,\,\beta},
  \blanc
 i=1,2,3, \lb{sp_vec}
\end{equation}
where $\sigma_i$ are Pauli matrices. In particular, the $3D$ spinor coordinates $\ell^{\alpha\beta}$
(restricted by the condition $(\ell^{\alpha\beta})^* = - \ell_{\alpha\beta}$) correspond to real vector coordinates $\ell_i$.  The only exception
from the rule \p{sp_vec}
is the relation between the partial derivatives $\partial_{\alpha\beta} = \partial/\partial \ell^{\alpha\beta}$ and
$\partial_i =  \partial/\partial \ell_i$,
\begin{equation}
\partial_{\alpha\beta}= -\frac{1}{2}\left(\sigma_i\right)_{\!\alpha\beta}\partial_i,\blanc
  \partial_i= -\left(\sigma_i\right)_{\!\alpha}^{\,\,\beta}\partial^\alpha_\beta\,.
\end{equation}
We also make a similar conversion of the gauge group indices,
\begin{equation}
 M_{\!\gamma}^{\,\,\delta}=\frac{1}{2} M^a \left(\sigma_a\right)_{\!\gamma}^{\,\,\delta},
  \blanc
 M^a=M_{\!\delta}^{\,\,\gamma}\left(\sigma_a\right)_{\!\gamma}^{\,\,\delta},
  \blanc
 a=1,2,3\,,
\end{equation}
for any Hermitian traceless $2\times 2$ matrix $M$, and define
\begin{equation}
 T^a= \frac{1}{2}\bar\vp^\alpha \left(\sigma_a\right)_{\!\alpha}^{\,\,\beta}\vp_\beta\,.\lb{DefT}
\end{equation}

In the new notations, the total Lagrangian \p{ACT} takes the following form:
\begin{multline}\label{vectLagr}
{\cal L}=
  \frac 12 f^{-2}\dot \ell_i^2
  +\A_i^a T^a\dot \ell_i
  +i\bar\vp^\alpha\left(\dot\vp_\alpha+iB\vp_\alpha\right)+kB
  +i\bar\psi_\alpha\dot\psi^\alpha
  +f^2\nabla_i U^a T^a\,\psi\sigma_i\bar\psi
\\[2mm]
  +\frac{1}{4}\left\{f\partial_i^2 f-3\left(\partial_i f\right)^2\right\}\psi^4
  +2 f^{-1}\varepsilon_{ijk}\partial_i f\,\dot\ell_j\, \psi\sigma_k\bar\psi
\\[2mm]
 +\frac 18f^{-2}F^2
  +\frac 12 F\left( U^a T^a
- f^{-1}\partial_i f\,\psi\sigma_i\bar\psi\right).
\end{multline}
Here
\begin{equation}
\nabla_i U^a=\partial_i U^a +\varepsilon^{abc}\A_i^b U^c
\end{equation}
and the Bogomolny equations \p{samo} relating $\A_i^a$ and $U^a$ are equivalently
rewritten in the more familiar form,
\begin{equation}
{\cal F}_{ij}^a = \varepsilon_{ijk}\nabla_k U^a\,, \lb{samoV}
\end{equation}
where ${\cal F}_{ij}^a = \partial_i\A_j^a - \partial_j\A_i^a
+\varepsilon^{abc}\A_i^b\A_j^c$.
Finally, the gauge field and the matrix potential defined in \p{GpotMpot} are rewritten as
\begin{equation}\label{3dsamo}
 \A_i^a
  =-\varepsilon_{ija} \partial_j \ln h,\blanc
  U^a=-\partial_a \ln h\,, \qquad \Delta\,h = 0\,.
\end{equation}

The component action corresponding to the Lagrangian \p{vectLagr} is partly on shell
since we have already eliminated the fermionic fields of the auxiliary $v^+$
multiplet by their algebraic equations of motion. The fields of the coordinate multiplet $L^{++}$
are still off shell.
The ${\cal N}=4$ transformations leaving invariant the action $S = \int dt \,{\cal L}$ look most transparent in terms of the
component fields $\ell_i, F, \chi^\alpha$, $\bar\chi^\alpha$, $\phi^\beta$, $\bar\phi^\beta$:
\begin{equation}
\begin{array}{lll}
&\delta \ell_i = -\frac{i}{2}\left(\epsilon\sigma_i\chi + \bar\epsilon\sigma_i\bar\chi \right)\,,
&\delta F = \epsilon^\alpha\dot{\chi}_\alpha + \bar\epsilon^\alpha\dot{\bar\chi}_\alpha\,,
\\
&\delta \chi^\alpha = iF\bar\epsilon^\alpha + 2(\bar\epsilon\sigma_i)^\alpha\,\dot{\ell}_i\,,
&\delta \bar\chi^\alpha = -iF\epsilon^\alpha - 2(\epsilon\sigma_i)^\alpha\,\dot{\ell}_i\,,
\\
&\delta \phi^\alpha = \frac{i}{2}\left(\epsilon^\alpha \chi\sigma_i\phi
+ \bar\epsilon^\alpha \bar\chi\sigma_i\phi\right)\partial_i\ln h\,, \quad
&\delta \bar\phi^\alpha = \frac{i}{2}\left(\epsilon^\alpha \chi\sigma_i\bar\phi
+ \bar\epsilon^\alpha \bar\chi\sigma_i\bar\phi\right)\partial_i\ln h\,.
\end{array}
\lb{Transf}
\end{equation}
These transformations can be deduced from the analytic subspace realization of ${\cal N}=4$ supersymmetry \p{N4transf},
with taking into account the compensating $\U(1)$ gauge transformations of the superfields $v^+, \widetilde{v}^+$ and $V^{++}$
needed to preserve the WZ gauge \p{VPP}. Note that $\delta B = 0$ under ${\cal N}=4$ supersymmetry.\footnote{This transformation law
matches with the ${\cal N}=4, d=1$ superalgebra in WZ gauge, taking into account that the $d=1$ translation of $B$ looks
as a particular U(1) gauge transformation of the latter.}

After eliminating the auxiliary field $F$ by its equation of motion,
\be
F = 2 f^2\left( f^{-1}\partial_i f \,\psi\sigma_i \bar\psi - U^aT^a\right)\,, \lb{EqF}
\ee
the Lagrangian \p{vectLagr} takes the form
\begin{eqnarray}
{\cal L} &=&
  \frac 12 f^{-2}\dot \ell_i^2
  +\A_i^a T^a\dot \ell_i
  +i\bar\vp^\alpha\left(\dot\vp_\alpha+iB\vp_\alpha\right)+kB
  +i\bar\psi_\alpha\dot\psi^\alpha
  +f^2\psi\sigma_i\bar\psi\left(\nabla_i +f^{-1}\partial_i f\right) U^aT^a \nn
&&+\,\frac{1}{4}\left\{f\partial_i^2 f- 4\left(\partial_i f\right)^2\right\}\psi^4
  +2 f^{-1}\varepsilon_{ijk}\partial_i f\,\dot{\ell}_j\, \psi\sigma_k\bar\psi
 -\frac 12 f^2 ( U^a T^a)^2\,.\label{vectLagr1}
\end{eqnarray}
It is invariant, modulo a total time derivative,  under the transformations \p{Transf} in which $F$ is expressed from \p{EqF}.
We see that this Lagrangian involves three physical bosonic fields $\ell_i$ and
four physical fermionic fields $\psi_\alpha\,$. It is fully specified by two independent functions: the metric conformal factor $f(\ell)$ which
can bear an arbitrary dependence on $\ell_i$ and the function $h(\ell)$ which satisfies the $3D$ Laplace equation
and determines the background non-Abelian gauge and scalar potentials. The representation \p{hdef} for $h$ in terms
of the analytic function $K(\ell^{++}, u)$ yields in fact a general solution of the $3D$ Laplace equation \cite{HSS}.
The Lagrangian \p{vectLagr1} also contains the ``semi-dynamical'' spin variables $\varphi_\alpha, \bar\varphi^\alpha\,$,
the role of which is the same as in the $4D$ case \cite{Ivanov:2009tw}: after quantization they ensure that $T^a$ defined in \p{DefT}
become matrix SU(2) generators corresponding to the spin $|k|/2$ representation.

\section{Hamiltonian and supercharges}
The Lagrangian \p{vectLagr1} is the point of departure for setting up the Hamiltonian formulation of the model under consideration and
quantizing the latter.
The main peculiarity of the quantization procedure in the present  case is related to the spin variables $\varphi_\alpha,
\bar\varphi^\alpha\,$. The corresponding commutation relations are
\be
[\varphi_\alpha, \bar\varphi^\beta] = \delta^\alpha_\beta\,,  \quad \quad [\varphi_\alpha, \varphi_\beta]
= [\bar\varphi^\alpha, \bar\varphi^\beta] = 0\,,
\ee
whence, e.g., $\varphi_\alpha\rightarrow \hat\varphi_\alpha \equiv \partial/\partial \bar\varphi^\alpha\,$ and the constraint \p{Cconstr}
becomes the condition on the wave functions
\be
 \bar\varphi^\alpha\frac{\partial}{\partial \bar\varphi^\alpha}\Psi = k\Psi\,\lb{Qconstr}
\ee
(hereafter, without loss of generality, we assume that $k > 0\,$).
It implies that $\Psi$ is a collection of homogeneous monomials of $\bar\varphi^\alpha$ of an integer degree $k$ and, thus,
carries an irreducible $\SU(2)$ multiplet with spin $k/2$ (the number of such independent monomials is equal just to $k+1$).
The $\SU(2)$ vector $T^a$ defined in \p{DefT} satisfy the $\SU(2)$ commutation relations
\be
[T^a, T^b] = i\varepsilon^{abc}T^c\,,
\ee
and, as a consequence of the constraint \p{Qconstr}, is subject to the condition
\be
T^aT^a = \frac{k}{2}\left(\frac{k}{2} +1 \right).
\ee
In this way, $T^a$ can be treated as generators of the irreducible unitary representation of $\SU(2)$ with spin $k/2$.
\footnote{The crucial role of the constraint \p{Qconstr} is to restrict the space of quantum states of the considered model
to the {\it finite} set of irreducible $\SU(2)$ multiplets of fixed spins (e.g., of the spin $k/2$ in the bosonic sector).
This is an essential difference of our approach from that employed, e.g.,  in \cite{Bala} (and, lately, in \cite{BKS,KLS})
where no any analog of the constraints
\p{Cconstr} and \p{Qconstr} is imposed, thus allowing the space of states to involve an {\it infinite} number of $\SU(2)$ multiplets
of all spins. The quantization scheme which we follow here can be traced back to the work \cite{Polychronakos:1991bx}.
In the SQM context, it was already used in \cite{FIL} and \cite{Ivanov:2009tw}.}

The system (\ref{vectLagr1}) is a generalization, to the non-Abelian case,
of the Abelian ${\cal N}=4$ $3D$ system found in \cite{Smilga:1986rb}, which, in turn, is a generalization, to the conformal metric,
of the system in a flat space invented by
de Crombrugghe and Rittenberg \cite{de Crombrugghe:1982un}. After substitution
of $\SU(2)$ spin-$k/2$ generators instead of $T^a$ \cite{Ivanov:2009tw}, the (quantum) Hamiltonian of this system takes the form
\begin{multline}\label{decrom}
  H=\frac{1}{2}f \left(\hat p_i-\ca A_i\right)^2 f
+\frac 12 f^2 U^2
-f^2\nabla_i U\psi\sigma_i\bar\psi
\\[2mm]
+\Big\{
\ve_{ijk}f\partial_i f\left(\hat p_j-\A_j\right)
-f\partial_k f U
\Big\}\psi\sigma_k\bar\psi
    + f\partial^2 f\left\{\psi^{\gamma}\bar\psi_{\gamma}-\frac{1}{2}\left(\psi^{\gamma} \bar\psi_{\gamma}\right)^2\right\},
\end{multline}
which is just a static $3D$ reduction of the 4-dimensional Hamiltonian given in \cite{KonSmi}.
In this expression, the gauge field $\A_i = \A_i^aT^a$ and the scalar potential $U = U^aT^a$ are $\SU(2)$ matrices subjected
to the constraint (\ref{samoV}). It is also easy to find the supercharges $Q_\alpha, \bar Q^\beta$,
\begin{equation}\label{QQ}
\begin{array}{l}
  Q_\alpha = f \left(\sigma_i \bar\psi\right)_\alpha \left(\hat p_i-\ca A_i\right)
-\psi^{\gamma} \bar\psi_{\gamma} \left(\sigma_i\bar\psi\right)_\alpha i\partial_i f
-ifU\bar\psi_\alpha,
\\[2mm]
  \bar Q^\alpha = \left(\psi\sigma_i\right)^\alpha \left(\hat p_i-\ca A_i\right)f
+i\partial_i f \left(\psi\sigma_i\right)^\alpha \psi^{\gamma}\bar\psi_{\gamma}
+ifU\psi^\alpha,
\end{array}
\end{equation}
\begin{equation}
\{Q_\alpha, \bar Q^\beta \} = 2\delta^\beta_\alpha\,H\,, \quad \{Q_\alpha, Q_\beta \} = \{\bar Q^\alpha, \bar Q^\beta \} = 0\,.\lb{SUSY}
\end{equation}
The ordering ambiguity arising in the case of the general conformal factor $f(\ell)$ can be fixed, as in \cite{KonSmi},
by the arguments of Ref.~\cite{SMI}.

We would like to emphasize that the only condition required from the background matrix fields ${\cal A}_i$ and $U$ for the generators
$Q_\alpha$ and $\bar Q^\beta$ to form ${\cal N}=4$ superalgebra \p{SUSY} is that these fields satisfy the Bogomolny equations \p{samoV}.
Thus the expressions \p{decrom} and \p{QQ} define the ${\cal N}=4$ SQM model in the field of {\it arbitrary} BPS monopole,
not necessarily restricted to the ansatz \p{3dsamo}. Also, one can extend the gauge group $\SU(2)$ to $\SU(N)$ in \p{decrom} and \p{QQ}.
The 't Hooft type ansatz \p{3dsamo} and the choice of $\SU(2)$ as the gauge group are required for the existence
of {\it off-shell} Lagrangian formulation of this SQM system. We do not know whether the most general system can be derived from some
off-shell superfield formalism, though the corresponding component Lagrangian with the on-shell realization of ${\cal N}=4$ supersymmetry
can certainly be constructed. It is a straightforward extension of the Lagrangian \p{vectLagr} or (\ref{vectLagr1}), with the properly enlarged
set of semi-dynamical spin variables, and the external potentials ${\cal A}_i$, $U$ taking values in the $su(N)$ algebra and
obeying Eq.~\p{samoV}. This situation is quite similar to what was observed in \cite{KonSmi, Ivanov:2009tw}
in the case of $4D$ SQM with self-dual gauge fields.

Finally, as a simple example of the monopole background consistent with the off- and on-shell ${\cal N}=4$ supersymmetry,
let us consider a particular $3D$ spherically symmetric case. It corresponds
to the most general $\SO(3)$ invariant solution of the Laplace equation for the function $h$
\be
h_{{\rm so}(3)}(\ell) = c_0 + c_1\,\frac{1}{\sqrt{\ell^2}}\,. \lb{so3}
\ee
The corresponding potentials calculated according to Eqs.~\p{3dsamo} read
\be
{\cal A}^a_i= \varepsilon_{ija}\frac{\ell_j}{\ell^2} \,\frac{c_1}{c_1 + c_0\sqrt{\ell^2}}, \quad U^a =
\frac{\ell_a}{\ell^2} \,\frac{c_1}{c_1 + c_0\sqrt{\ell^2}}. \lb{WYa}
\ee
This configuration becomes the Wu-Yang monopole \cite{WYa} for the choice $c_0 = 0\,$.
It is easy to find the analytic function $K(\ell^{++},u)$
which generates the solution \p{so3} (see \cite{IvLecht}):
\be
&&  h_{{\rm so}(3)}(\ell) = \int du\, K_{{\rm so}(3)}(\ell^{++}, u)\,, \quad K_{so(3)}(\ell^{++}, u)
= c_0 + c_1 \left(1 + a^{--}\hat{\ell}^{++}\right)^{-\frac{3}{2}}\,, \lb{Kso3} \\
&& \ell^{++} \equiv \hat{\ell}^{++} + a^{++}\,, \quad
a^{\pm\pm} = a^{\alpha\beta}u^\pm_\alpha u^\pm_\beta\,, \quad a^{\alpha}_\beta a^\beta_\alpha = 2\,. \nonumber
\ee

One could equally choose as $h(\ell)$, e.g., the well-known multi-center solution to the Laplace equation, with the broken $\SO(3)$.
Note that the ${\cal N}=4$ mechanics with
coupling to Wu-Yang monopole was recently constructed in \cite{BKS}, proceeding from a different approach, with
the built-in $\SO(3)$ invariance and the treatment of spin variables in the spirit of Ref. \cite{Bala}.
Our general consideration shows, in particular, that the demand of $\SO(3)$ symmetry
is not necessary for the existence of ${\cal N}=4$ SQM models with non-Abelian
monopole backgrounds.

\section{Relation to four-dimensional ${\cal N}=4$ SQM model}

It is instructive to show that \p{3dsamo} can indeed be viewed as a $3D$ reduction of 't~Hooft ansatz for the solution
of general self-duality equation in $\mathbb{R}^4$ for the gauge group $\SU(2)$, with the identification $U^a =  \A_0^a\,$,
while the condition \p{samoV} as $3D$ reduction of this equation.

To establish this relation,
we use the following dictionary between the $\SO(4)\sim \SU(2)\times \SU(2)$ spinor formalism of Refs.~\cite{KonSmi,Ivanov:2009tw}
and its $\SU(2)$ reduction:
\begin{equation} \lb{dict}
\begin{array}{c}
\left(\sigma_\mu\right)_{\alpha\dot\beta}\rightarrow
  \left\{i\delta_\alpha^{\beta},\,
   \left(\sigma_i\right)_{\!\alpha}^{\,\,\beta}\right\}, \quad \varepsilon^{\dot\alpha\dot\beta} \rightarrow - \varepsilon_{\alpha\beta}\,,
   \varepsilon_{\dot\alpha\dot\beta} \rightarrow - \varepsilon^{\alpha\beta}\
\\[2mm]
  x_{\alpha\dot\beta}\rightarrow \ell_\alpha^{\beta}, \blanc x^{\alpha\dot\beta}\rightarrow -\ell^\alpha_{\beta}\blanc
  \psi_{\dot\alpha}\rightarrow\psi^{\alpha}.
\end{array}.
\end{equation}
This reflects the fact that the R-symmetry $\SU(2)$ in the $({\bf 3, 4, 1})$ models can be treated as a diagonal subgroup in the
symmetry group $\SO(4)$ $\sim  \SU(2)\times \SU(2)$ of the $({\bf 4, 4, 0})$ models, with the $\SU(2)$ factors acting, respectively,
on the undotted and dotted indices.

The self-dual $\mathbb{R}^4$ SU(2) gauge field in the 't Hooft ansatz used in \cite{Ivanov:2009tw} can be written in the spinor notation as
\begin{eqnarray}
&&({\cal A}_{\alpha\dot\rho})^{\,\,\gamma}_{\!\beta}  = -\frac{2i}{h}\left(\varepsilon_{\alpha\beta}\,\partial^\gamma_{\dot\rho}h -
\frac{1}{2}\,\delta^\gamma_\beta\, \partial_{\alpha\dot\rho}h \right), \qquad \partial_{\alpha\dot\rho} \equiv
\frac{\partial}{\partial x^{\alpha\dot\rho}}\,, \nn
&& h = h(x^{\alpha\dot\beta})\,, \quad
\partial^{\alpha\dot\beta}\partial_{\alpha\dot\beta}\, h = 0\,.\lb{sd4}
\end{eqnarray}
Then, using the rules \p{dict}, one performs the reduction $\mathbb{R}^4 \rightarrow \mathbb{R}^3$ as
\begin{equation}\label{Ared}
\begin{array}{l}
({\cal A}_{\alpha\dot\beta})_{\!\gamma}^{\,\,\delta}
\; \rightarrow \; iU_{\!\gamma}^{\,\,\delta} \delta_\alpha^\beta + ({\cal A}_{\alpha}^{\beta})_{\!\gamma}^{\,\,\delta}\,,\;\;\quad
({\cal A}_{\alpha}^{\alpha})_{\!\gamma}^{\,\,\delta} = 0\,,
\\[2mm]
h(x)\; \rightarrow \;h(\ell), \quad
\partial_\beta^\alpha\partial^\beta_\alpha\,h = 0\,.
\end{array}
\end{equation}

Upon this reduction, the four-dimensional ansatz \p{sd4} yields precisely \p{GpotMpot}, while the general self-duality condition
\be
2\partial_{\alpha\dot\rho}({\cal A}_{\beta}^{\dot\rho})^{\,\,\delta}_{\!\gamma} + i({\cal A}_{\alpha\dot\rho})^{\,\,\lambda}_{\!\gamma}
({\cal A}_{\beta}^{\dot\rho})^{\,\,\delta}_{\!\lambda} + (\alpha \leftrightarrow \beta) = 0 \lb{sd41}
\ee
goes over into the Bogomolny equations \p{samo}. Of course, the same reduction can be performed in the vector notation,
with
$\F_{\mu\nu}\rightarrow\left\{\F_{ij}, \F_{0k}=\nabla_k U\right\}$,
and Eqs.~\p{samoV}, \p{3dsamo} as an output.

Thus, the general gauge field background prescribed by the off-shell ${\cal N}=4$
supersymmetry in our  $({\bf 3,4,1})$ model is a static form of the 't Hooft ansatz
for the self-dual $\SU(2)$ gauge field in $\mathbb{R}^4\,$. As was shown in \cite{Ivanov:2009tw}, this particular form of the self-dual field
is prescribed by the same off-shell ${\cal N}=4$ supersymmetry in the 4D SQM model based on the supermultiplet $({\bf 4,4,0})$.
This suggests that the above bosonic target space reduction has its superfield counterpart relating the model of \cite{Ivanov:2009tw}
to the one considered in the present paper.

Indeed, the superfield $({\bf 3,4,1})$ action \p{ACT} can be obtained from
the $({\bf 4,4,0})$ multiplet action of Ref.~\cite{Ivanov:2009tw} via the ``automorphic duality'' \cite{GR} by considering a restricted
class of the $({\bf 4,4,0})$ actions
with $\U(1)$ isometry and performing a superfield gauging of this isometry by an extra gauge superfield $V^{++}{}'$ along the general
line of Ref.~\cite{DI}. Actually, the bosonic target space reduction we have just described corresponds to the shift isometry
of the analytic superfield $q^{+\dot\alpha}$ accommodating the $({\bf 4,4,0})$ multiplet, namely,
to $q^{+\dot\alpha} \rightarrow q^{+\dot\alpha} + \omega u^{+ \dot\alpha}\,$. It is the invariant projection
$q^{+\dot\alpha}u^+_{\dot\alpha}$ which is going to become the $({\bf 3,4,1})$ superfield $L^{++}$ upon gauging
this isometry and choosing the appropriate manifestly ${\cal N}=4$ supersymmetric gauge. Another type of possible
isometry of the $q^{+\dot\alpha}$ actions of Ref.~\cite{Ivanov:2009tw} is the phase one, with $q^{+\dot{1}}q^{+\dot{2}}\,$
as the appropriate invariant. It can also be gauged, with the same $L^{++}$ action as a result.

An important impact of this superfield reduction on the structure of the component action is the appearance
of the new induced potential bilinear in the gauge group generators $\sim U^2 = U^a U^b T^a T^b\,$. It comes out
as a result of eliminating the auxiliary field $F$ in the off-shell $({\bf 3, 4, 1})$ multiplet, and so
is necessarily prescribed by ${\cal N}=4$ supersymmetry. It is interesting that analogous potential
terms were introduced in \cite{H2} at the bosonic level for ensuring the existence of some hidden
symmetries in the models of $3D$ particle in a non-Abelian monopole background.

The same reduction ${\mathbb R}^4\rightarrow {\mathbb R}^3$ can be performed
at the level of Hamiltonian and supercharges. In particular, the reduction
of the Hamiltonian of the $4D$ system of \cite{KonSmi} yields the $3D$ Hamiltonian~(\ref{decrom}).

\section{Conclusions}
In this paper, we constructed some rather general off-shell ${\cal N}=4$ supersymmetric coupling of the $d=1$ multiplet ${\bf (3,4,1)}$ to an
external $\SU(2)$ gauge field. The off-shell ${\cal N}=4$ supersymmetry restricts the latter to be a $3D$ reduction of the 't Hooft ansatz
for self-dual $\SU(2)$ gauge field in $\mathbb{R}^4$, that is a particular solution of the Bogomolny monopole equations.
At the component level, the coupling to a gauge field is necessarily accompanied
by an induced potential which is bilinear in the $\SU(2)$ generators and arises as a result of eliminating an auxiliary field.
Our main devices, as in \cite{Ivanov:2009tw}, were the HSS approach and the use of an analytic ``semi-dynamical'' multiplet $({\bf 4,4,0})$
with the WZ type action. This multiplet incorporates  $\SU(2)$ doublet bosonic spin variables which are crucial for arranging
couplings to non-Abelian gauge fields. We also presented the explicit form of the corresponding Hamiltonian and ${\cal N}=4$
supercharges which can be equally used for an arbitrary monopole BPS background, though with the on-shell realization
of ${\cal N}=4$ supersymmetry.

Like in the case of $4D,$ ${\cal N}=4$ mechanics coupled to a self-dual non-Abelian gauge field \cite{Ivanov:2009tw},
in the $3D$ case considered here there remains a problem of extending the model to a generic $\SU(N)$
gauge group, as well as to general monopole backgrounds obtained as a $3D$ reduction of ADHM construction \cite{Atiyah:1978ri}.
It would be also interesting to study SQM models with nonlinear counterparts of the target space multiplet $({\bf 3, 4, 1})$ \cite{IvLecht,IKLecht}
and/or of the semi-dynamical multiplet $({\bf 4, 4, 0})$ \cite{DI}. Such models exhibit more general target geometries as compared to
the conformally-flat ones associated with the linear $({\bf 3, 4, 1})$ multiplet and are capable
to yield also more general background gauge fields.

Finally, it is worthwhile to note that similar constraints (Bogomolny equations) on the external non-Abelian $3D$ gauge field were
found in \cite{Berry}, while considering an ${\cal N}=4$ extension of Berry phase in quantum mechanics. However, no invariant actions
and/or the explicit expressions for the Hamiltonian and ${\cal N}=4$ supercharges were presented there.

\section*{Acknowledgments}
We cordially thank Andrei Smilga for numerous useful discussions and comments. E.I. thanks the Directorate of SUBATECH, Nantes,
where this work was completed, for warm hospitality. His work was supported in part by the RFBR grants 09-02-01209, 09-02-91349
and 09-01-93107.

\end{document}